\documentclass[11pt]{article}

\usepackage{jheppub}
\usepackage[T1]{fontenc}
\usepackage[latin9]{inputenc}
\usepackage{amsmath}
\usepackage{amssymb}
\usepackage{setspace}
\usepackage{float}\usepackage{amstext}
\makeatother

\begin{document}

\title{A Non-Renormalization Theorem in Gapped Quantum Field Theory}
\author{Tomer Shacham}
\emailAdd{tomer.shacham@phys.huji.ac.il}
\affiliation{Racah Institute of Physics, \\ The Hebrew University of Jerusalem, \\ Jerusalem, 91904, Israel}

\abstract{
We discuss the two-point functions of 
the U(1) current and energy-momentum tensor in certain gapped three-dimensional field theories, and show that the parity-odd part in both of these correlation functions is one-loop exact. In particular, we find a new and simplified derivation of the Coleman-Hill theorem that also clarifies several subtleties in the original argument. For the energy momentum tensor, our result means that the gravitational Chern-Simons term for the background metric does not receive quantum corrections.
}

\maketitle

\def \b{\begin{equation}}
\def\e{\end{equation}}
\def \O{\mathcal{O}}
\def \F{\mathcal{F}}
\def \P{\mathcal{P}}
\def \A{\mathcal{A}}
\def \B{\mathcal{B}}
\def \<{\langle}
\def \>{\rangle}
\def \x{{\scriptstyle (x)}}
\def \y{{\scriptstyle (y)}}
\def \z{{\scriptstyle (z)}}
\def \p{{\scriptstyle (p)}}
\def \q{{\scriptstyle (q)}}
\def \0{{\scriptstyle (0)}}
\def \mp{{\scriptstyle (-p)}}
\newcommand{\K}[1]{{\scriptstyle (k_{#1})}}
\newcommand{\OO}[1]{{ \O({\rm momentum}^{#1})}}

\section{Introduction}
In quantum field theories, the classical description is often modified by quantum corrections; in some cases, these corrections are limited by non-renormalization theorems. 
In %the context of 
2+1 dimensional gauge theories \citep{Siegel1979,Jackiw1981,Deser1982,Schonfeld1981} with 
massive matter, it was found that the topological mass of the gauge field receives no contributions at two loops \citep{Kao1985,Bernstein1985}.\footnote{The topological mass contributes a parity-odd term to the gauge field propagator. At zero momentum, this term is the Chern-Simons coefficient in the effective action.}

Following this discovery, Coleman and Hill proved that no such corrections are possible when the 
 gauge group is  Abelian \citep{Coleman1985}. Their proof deals directly with properties of Feynman graphs. They explicitly show that above one loop, the contribution of each subgroup of graphs with similar topologies is $\OO{2}$ and therefore cannot modify the low energy effective action.
 The reason behind the difference between one and many loops remained implicit in the proof. 

Here, we re-derive the Coleman-Hill theorem as a specific case of a more general theorem, and provide a natural explanation for the distinction of the one loop graph. %As we show, the  one loop contribution  is in some sense classical, and therefore stands out in front of the perturbative expansion. 
We then prove a closely related statement regarding the non-renormalization of the parity-odd part of $\<T_{\mu\nu} T_{\rho\sigma}\>$ at zero momentum.

Our argument is simple. Since in Abelian theories  $A_\mu$ itself is neutral, the parity-odd part (at zero momentum) of $\<A_\mu A_\nu\>$ is the same as for a background gauge field which couples to a global $U(1)$ current, $j_\mu$. We argue that if  $\<j_\mu j_\nu O\>$ is $\OO{2}$  for an arbitrary scalar $O$, so is the rest of the perturbative expansion. As it turns out, this must be the case for $\<j_\mu j_\nu O\>$ to satisfy the corresponding Ward identity. Finally, we trace the reason behind this non-renormalization to the fact that the Chern-Simons Lagrangian is not gauge invariant.

This paper is organised as follows. In section \ref{section:Conditions} we specify the class of theories for which the theorem applies. In section \ref{section:WardIdentities} we briefly review Ward identities in order to establish notation. Sections \ref{section:U1} and \ref{section:EM} present the examples of the $U(1)$ current and energy-momentum tensor, respectively. Finally, in section \ref{section:Discussion}, we conclude.

\section{Conditions}
\label{section:Conditions}
We are interested in theories described by an action $\mathcal{S}=\int\!\!\mathcal{L}$  with
\b
\mathcal{L}=\mathcal{L}_{0}+\sum_{i}\lambda_{i}O_{i},
\label{eq:Coonditions}
\e
where $\lambda_{i}$ are couplings and $O_{i}$ are scalar operators.\footnote{
The subscript $_0$ will henceforth  denote the limit $\lambda_{i}\!\!\rightarrow\!0$, which
corresponds to the free theory.} 

We further restrict the discussion to cases where $\mathcal{L}_{0}$ describes a massive theory which admits to a manifestly Lorentz invariant quantization.  In particular, we study three dimensional Abelian gauge theories, 
QED$_3$, with arbitrary massive matter.
For this class of theories,
\b
\mathcal{L}_{0}=\mathcal{L}_{\rm gauge}+\mathcal{L}_{\rm matter}
\e
where
\b
\mathcal{L}_{\rm gauge}=\frac{1}{4}F_{\mu\nu}\left(\frac{1}{e^{2}}F^{\mu\nu}-i\kappa\epsilon^{\mu\nu\rho}A_{\rho}\right)-\frac{1}{2\xi e^{2}}\left(\partial_{\mu}A^{\mu}\right)^{2}
\e
 describes a gauge field with mass $\kappa e^{2}$.\footnote{Throughout this paper, we use a Euclidean metric.} 

\section{A brief review of Ward identities}
\label{section:WardIdentities}
When a theory is invariant under a continuous global transformation,
Noether's theorem guarantees that there exists a classically conserved current, such that
\begin{equation}
\frac{\delta}{\delta\epsilon\x}\mathcal{S}'+\partial^{\mu}j_{\mu}\x=0,
\label{eq:NoetherCurrent}
\end{equation}
where $\epsilon$ is the continuous parameter of the transformation and $'$ denotes a transformed object. 
If the symmetry is not anomalous (i.e. it survives quantization), correlation functions are
independent of the variation.\footnote{Given an observable operator $\A$,
% i.e. a functional of the fields $\Phi$ and their derivatives, 
we define the correlation function by $\<\A\>\equiv\int\!\mathcal{D}\Phi e^{-\mathcal{S}}\A$.}  In particular,
\begin{equation}
\frac{\delta}{\delta\epsilon\x}\left\langle j_{\nu}\y O\z \right\rangle '=0,
\end{equation}
and so
\b
 \frac{\partial}{\partial x_{\mu}}\< j_{\mu}\x j_{\nu}\y O\z\> = -
  \< \frac{\delta}{\delta\epsilon\x}j'_{\nu}\y O\z\> -\< j_{\nu}\y\frac{\delta}{\delta\epsilon\x}O'\z\rangle.
\label{eq:GenericWardIdentityXspace} 
\e
We refer to (\ref{eq:GenericWardIdentityXspace}) as the Ward identity 
for $\langle j_{\mu}j_{\nu}O\rangle$.

\section{The $U(1)$ current}
\label{section:U1}

Here we take $j_{\mu}$ to be the current associated with a global $U(1)$ symmetry, which acts on the charged fields as 
\b
\Phi\rightarrow\Phi'=e^{ie\epsilon}\Phi,
\e
with $e$ the ``electric'' charge. 

In a massive theory there are no infrared singularities, and so the most general tensor structure, or parameterization, of $\< j_{\mu} \p j_{\nu} \mp \>$ reads
\b
a\delta_{\mu\nu}+\delta\kappa\, \epsilon_{\mu\nu\rho}p^{\rho}+\OO{2}.
\label{eq:JJparameterization}
\e
Note that only terms in the perturbative expansion of $\< j_{\mu}\p j_{\nu}\mp\>$ that are $\OO{}$ can possibly contribute to $\delta\kappa$.\footnote{$\delta\kappa$ is projected out by \b
\lim_{p\rightarrow0}\left(\frac{1}{6}\epsilon^{\mu\nu\lambda}\frac{\partial}{\partial p^{\lambda}}\< j_{\mu}\p j_{\nu}\mp\>\!\right).
\e} As advocated, we argue that no such quantum corrections exist. 

The basic observation is that the entire
 perturbative expansion
around the free theory consists of zero momentum insertions of the scalar operators $O_{i}$:
\b
 \< j_{\mu}\p j_{\nu}\mp\>  =\< j_{\mu}\p j_{\nu}\mp\>_{0}
-\sum_{i}\lambda_{i}\< j_{\mu}\p j_{\nu}\mp O_{i}\0\>_{0}
  +\sum_{ij}\frac{1}{2}\lambda_{i}\lambda_{j}\<j_{\mu}\p j_{\nu}\mp O_{i}\0 O_{j}\0 \>_{0}+...\ .
\e

As the theory is gapped $\< j_{\mu}\p j_{\nu}\mp O_1\0...O_n\0\>$ is well defined as the limit
\b
{\rm lim}_{k_{i}\rightarrow0}\< j_{\mu}\p j_{\nu}\q O_{1}\K{1}...O_{n}\K{n}\>,
 \label{eq:<ABOOO>}
\e
where $q=-p-\Sigma_{i}k_{i}$. We can therefore take this limit in two steps, $k_{i\neq1}\!\rightarrow\!0$ followed by $k_1\!\rightarrow\!0$.

Consider the parameterization of 
$\<j_{\mu}\p j_{\nu}\q O_{1}\K{1}O_{2}\0...O_{n}\0\!\>$. 
The insertion of $O_{1}\K{1}$ allows $p$ and $q$ to be independent;
the insertions at zero momentum, on the other hand,  do not impose or relax any constraints
on the tensor structure.\footnote{Note that relying
on the tensor structure is permissible only when the quantization is manifestly Lorentz invariant.
} Consequently, this parameterization does not depend on the number of insertions at zero momentum; the different orders in perturbation theory may differ only by the
coefficients in the momentum expansion.  Therefore, if $\O \! \left({\rm momentum}\right)$ terms in the parameterization of $\<j_{\mu}\p j_{\nu}\q O \K{1}\>$ violate the corresponding Ward identity, they are absent from the rest of the perturbative corrections as well.
As we now show, this is indeed the case. %Even though the symmetry group is Abelian,  

For every charged scalar $\phi$,  the variation of the current picks up a term
\b
\frac{\delta}{\delta\epsilon\x}\left(\!i\bar{\phi}\overleftrightarrow{\partial_{\mu}}\phi\!\right)'\!\!\!\y=-2e\bar{\phi}\phi\y\partial_{\mu}\delta{\scriptstyle\left(x-y\right)},
\e
and the Ward identity for $\< j_\mu j_\nu O\>$ reads
\b
\frac{\partial}{\partial x_{\mu}}\Big(\! \< j_{\mu}\x j_{\nu}\y O\z\>-2e\delta_{\mu\nu}\delta{\scriptstyle \left(x-y\right)} \< \bar{\phi}\phi\y O\z\> \! \Big)=0.
\e
The existence of this contact term is due to the absence of
the seagull term, as the symmetry is global.
However, correlation functions are defined up to contact terms;  as $\delta_{\mu\nu}\delta{\scriptstyle (x-y)}$ respects the symmetry 
of $j_{\mu}\x j_{\nu}\y$, one can absorb the contact term into a redefinition of $\langle j_\mu j_\nu O\rangle$. Hence, there exists a regularization scheme in which 
\b
p^{\mu}\< j_{\mu}\p j_{\nu}\q O{\scriptstyle(-p-q)}\> =0.
\label{eq:AbelianWardMomentum}
\e

The parameterization of  $\< j_{\mu}\p j_{\nu}\q O \K{1} \>$ is given by  
\b
a'\delta_{\mu\nu}+b\, \epsilon_{\mu\nu\rho}\left(p^{\rho}-q^{\rho}\right)+\OO{2}.
\label{eq:MomentumExpansion<JJO>}
\e
Since the Ward identity must be satisfied at each order in the momentum expansion, imposing (\ref{eq:AbelianWardMomentum}) on the parameterization of $\< j_{\mu}\p j_{\nu}\mp\>$ gives $a=0$. Similarly, one finds that both $a'$ and $b$ vanish;  $\< j_{\mu}j_{\nu}O\>$ is then $\OO{2}$ for an arbitrary  scalar $O$. 
We thus conclude that $\delta\kappa$ is completely determined by $\< j_{\mu}\p j_{\nu}\mp\>_0$ and does not receive perturbative corrections. As the current is quadratic
in the fields, the classical contribution corresponds to a one loop graph. 

The reason why the Ward identity prohibits $\O\!\left({\rm momentum}\right)$ terms in  $\<j_\mu j_\nu O\>$ can be understood by coupling the global $U(1)$ current to a background gauge field $a_\mu$, and the deformation $O_i$ to a background source $J_i$.\footnote{In the language of background fields, the Ward identity corresponds to  the invariance of the generating functional under small gauge transformations.} One then defines 
\b
\< j_\mu j_\nu O\> \equiv \frac{\delta}{\delta a^\mu}\frac{\delta}{\delta a^\nu}\frac{\delta}{\delta J}\mathcal{Z}\left[a , J_i \right]\Big|_{a=0,J_i=0}
\e 
where
\b
e^{-\mathcal{Z}\left[a , J_i \right]}=\Big\< \! e^{-\!\int \!  \left(j^\mu a_\mu +\sum_i \! \lambda _i O_i J_i\right)}\!\Big\>.
\e

The only term in the derivative expansion of $\mathcal{Z}\left[a , J_i \right]$ with two $a$'s and a $J$ that includes only one derivative is 
\b
\int \! \! d^3 x J \epsilon^{\mu\nu\rho}a_\mu \partial_\nu a_\rho.
\label{eq:JCS}
\e  
As the Chern-Simons Lagrangian varies by a full derivative under a gauge transformation, (\ref{eq:JCS}) is gauge invariant only for a constant $J$. However, the limit of constant $J$ corresponds to $\lim_{p\rightarrow0}O\p$. As the theory is gapped, there are no singularities at zero momentum and so this limit is regular. Therefore, terms which can contribute to $\< j_\mu j_\nu O\>$ must  be of a higher order in the derivative expansion.

Let us now briefly sketch the original proof of Coleman and Hill. The basic argument is that subgroups of graphs which contain a propagator of a neutral field cannot contribute to 
\b
\lim_{p\rightarrow0}\left(\frac{1}{6}\epsilon^{\mu\nu\lambda}\frac{\partial}{\partial p^{\lambda}}\< A_{\mu}\p A_{\nu}\mp\>\!\right).
\e
A subsequent analysis of the various topologies, 
excluding the one-loop case, reveals that each subgroup is equivalent to a linear combination of graphs with neutral propagators, and thus do not contribute as well. 

We note that since the gauge field itself is neutral, the parity-odd part of its two point function is the same as for a background field; 
the previous discussion of the one loop exactness of the global $U(1)$ current captures the same physics and completes the proof.  In the language of currents, the one loop contribution is in some
sense classical, hence the difference between one and many loops.

\section{The energy-momentum tensor}
\label{section:EM}
Invariance of a theory under
the Poincar\'{e} group action 
\b
x^{\mu}\rightarrow x'^{\mu}=x^{\mu}+\epsilon^{\mu},
\label{eq:Diffeo}
\e
implies  the conservation of the energy-momentum tensor,
\b
\frac{\delta}{\delta\epsilon^{\mu}\x}\mathcal{S}'+\partial^{\nu}T_{\mu\nu}\x=0.
\label{eqEMdefinition}
\e
Note that (\ref{eqEMdefinition}) defines an equivalence class; we shall henceforth refer to $T_{\mu\nu}$ as the symmetric tensor, which is associated with the coupling to gravity.  

In this section, we go through the same procedure
of constraining the parameterizations of $\< T_{\mu\nu}O\>$, $\< T_{\mu\nu}T_{\rho\sigma}\>$, and $\< T_{\mu\nu}T_{\rho\sigma}O\>$  by the corresponding Ward identities, which we now derive for completeness.

Under the transformation (\ref{eq:Diffeo}), the fields vary by a Lie derivative with respect~to~$\epsilon$: 
\b
\Phi'=\Phi+L_{\epsilon}\Phi.
\label{eq:FieldTransformationEM}
\e
The variation of a scalar field $\phi$ is $(\epsilon\!\cdot\!\partial)\phi$,
%\b
%\delta\phi=(\epsilon\!\cdot\!\partial)\phi,
%\e
and so the Ward identity for $\langle T_{\mu\nu}O\rangle$ reads 
\b
p^\mu \< T_{\mu\nu}\p O\q \>=(p+q)_{\nu}\< O{\scriptstyle (p+q)} \>=0,
\label{eq<TO>}
\e
where the last equality follows from conservation of momentum.

Since $T_{\mu\nu}$  is symmetric, the parameterization of $\< T_{\mu\nu}\p O\mp \>$ is proportional to
\b
p_\mu p_\nu-p^2 \delta_{\mu\nu}+\mathcal{O}\left(p^4\right);
\label{eq:WardMomentum<TO>}
\e
taking $O$ to be the unit operator, one finds that
\b
\< T_{\mu\nu}\0\>=0.
\label{eq:<TO>}
\e 
For a symmetric rank 2 tensor, 
\begin{eqnarray}
\delta T_{\rho\sigma}&&=(\epsilon\!\cdot\!\partial)T_{\rho\sigma}+\big(
T_{\rho\nu}\partial_{\sigma}\epsilon^{\nu} +(\rho\!\leftrightarrow\!\sigma)\big)\nonumber\\
&&=(\epsilon\!\cdot\!\partial)T_{\rho\sigma}
+ \left(\Sigma_{\mu\nu,\rho\sigma}^{(-)\alpha\beta}+\Sigma_{\mu\nu,\rho\sigma}^{(+)\alpha\beta}\right)T_{\alpha\beta}\partial^\mu\epsilon^\nu,
\end{eqnarray}
where the second term was decomposed into $\mu\!\leftrightarrow\!\nu$ symmetric and antisymmetric components
\b
\Sigma_{\mu\nu,\rho\sigma}^{(\!\pm\!)\alpha\beta}\equiv\frac{1}{2}\delta_\rho^\beta\big(\delta_{\mu\sigma}\delta_{\nu}^{\alpha}\pm(\mu\!\leftrightarrow\!\nu)\big)+ (\rho\!\leftrightarrow\!\sigma).
\e
The Ward identity for $\< T_{\mu\nu}T_{\rho\sigma}O\>$ is then given by
 \begin{eqnarray}
&& \frac{\partial}{\partial x_{\mu}} \Big(\!  \< T_{\mu\nu}\x T_{\rho\sigma}\y O\z \> 
+\delta{\scriptstyle (x-y)} \left(\Sigma_{\mu\nu,\rho\sigma}^{(-)\alpha\beta}+\Sigma_{\mu\nu,\rho\sigma}^{(+)\alpha\beta}\right)\langle T_{\alpha\beta}\y O\z \rangle \!  \Big)\nonumber  \\
&& =-\delta{\scriptstyle (x-y)}\frac{\partial}{\partial x^{\nu}}\< T_{\rho\sigma}\x O\z\> -\delta{\scriptstyle (x-y)}\frac{\partial}{\partial x^{\nu}}\< T_{\rho\sigma}\y O\x\> . \label{eq:WardPosition<TTO>}
\end{eqnarray}
The contact term containing $\Sigma^{(+)}$ can be absorbed into
a redefinition of $\langle T_{\mu\nu}T_{\rho\sigma}O\rangle $
as it is symmetric in $\mu\leftrightarrow\nu$.\footnote{This redefinition is equivalent to taking $\epsilon_\mu$ to be a Killing vector, i.e.
\b
\partial_{\mu}\epsilon_{\nu}+\partial_{\nu}\epsilon_{\mu}=0.
\label{eq:KillingVector}
\e
} 
In momentum space, this reads
\begin{eqnarray}
p^{\mu}  \< T_{\mu\nu}\p T_{\rho\sigma}\q O{\scriptstyle (-p-q)}\>=&&-p^{\mu}\Sigma_{\mu\nu,\rho\sigma}^{(-)\alpha\beta}\left\langle T_{\alpha\beta}{\scriptstyle (p+q)}O{\scriptstyle (-p-q)}\right\rangle  +q_{\nu} \< T_{\rho\sigma} \q O{\scriptstyle (-q)}\> \nonumber\\   &&-\left(q+p\right)_{\nu}\< T_{\rho\sigma}{\scriptstyle (p+q)}O{\scriptstyle (-p-q)}\>.
 \label{eq:WardMomentum<TTO>}
\end{eqnarray}
Taking $O$ to be the unit operator and using (\ref{eq:<TO>}) gives 
\b
 p^{\mu}\< T_{\mu\nu}\p T_{\rho\sigma}\mp\>=0.
 \e
The parameterization of $\< T_{\mu\nu}\p T_{\rho\sigma}\mp\>$ is then constrained to be
\b
...+\delta\kappa_g \Big(\big(\epsilon_{\mu\rho\lambda}p^{\lambda }(p_\nu p_\sigma -p^2 \delta_{\nu\sigma})+(\mu \! \leftrightarrow \! \nu)\big) +\rho \! \leftrightarrow \! \sigma)\Big)+\OO{4}
\label{eq:<TT>}
\e
where the ellipsis stands for a momentum$^2$ piece, which plays no role in this discussion; only
$\OO{3}$ terms in the perturbative expansion can contribute to $\delta\kappa_g$.\footnote{
$\delta\kappa_g$ is projected out by %$\lim_{p\rightarrow0}\left(\frac{1}{90}\epsilon^{\mu\rho\lambda}\frac{\partial}{\partial p^{\lambda}}\! \frac{\partial}{\partial p_{\nu}}\! \frac{\partial}{\partial p_{\sigma}}\!\< T_{\mu\nu}\p T_{\rho\sigma}\mp\>\!\right).
\b
 \lim_{p\rightarrow0}\left(\frac{1}{150}\epsilon^{\mu\rho\lambda}\frac{\partial}{\partial p^{\lambda}}\! \frac{\partial}{\partial p_{\nu}}\! \frac{\partial}{\partial p_{\sigma}}\!\< T_{\mu\nu}\p T_{\rho\sigma}\mp\>\!\right).
%%\label{eq:OneLoopExactEM}
\e} 
 However, such terms in the parameterization of $\< T_{\mu\nu}\p T_{\rho\sigma}\q O\K{1}\>$,
\begin{eqnarray}
\Big(\!\!\!\!&&\big(c_{1}\left(p_{\nu}q_{\sigma}+q_{\nu}p_{\sigma}\right)+
c_{2}\left(p_{\nu}p_{\sigma}+q_{\nu}q_{\sigma}\right)+c_3\delta_{\nu\sigma}(p^2+q^2)+c_4\delta_{\nu\sigma}(p \cdot q) \big)\nonumber\\ 
&&\times  \epsilon_{\mu\rho \lambda}\left(p^{\lambda}-q^{\lambda}\right)+(\mu \! \leftrightarrow \! \nu) \Big)+(\rho \! \leftrightarrow \! \sigma),
\end{eqnarray}
cannot satisfy the Ward identity (\ref{eq:WardMomentum<TTO>}) as there are no momentum$^3$ terms in $\<T_{\rho\sigma} O\>$. 
By repeating the arguments of the previous section, we conclude that $\delta\kappa_g$  is completely determined by $\< T_{\mu\nu}\p T_{\rho\sigma}\mp\>_0$ and does not receive perturbative corrections.\footnote{ 
Considering the Noether current associated with the transformation 
(\ref{eq:FieldTransformationEM})
\b
\epsilon^\nu T_{\mu\nu}=\frac{\partial\mathcal{L}}{\partial\left(\partial_{\mu}\Phi\right)}L_{\epsilon}\Phi
-\epsilon_\mu\mathcal{L},
\label{eq:NoetherCurrentEM}
\e
one finds that only the second term in (\ref{eq:NoetherCurrentEM}) may not be quadratic in the fields, and it clearly cannot contribute to $\delta\kappa_g$. The classical contribution %(\ref{eq:OneLoopExactEM})%  
therefore corresponds to a one-loop graph.}

As in the case of the $U(1)$ current discussed in the previous section, the underlying reason for the non-renormalization of the stress tensor's two-point function stems from the symmetries of the generating functional. To see this, couple %the deformation $O_i$ to a background source $J_i$ and 
the energy-momentum tensor to a background metric $g_{\mu\nu}$~and~define
\b
\< T_{\mu\nu} \x T_{\rho\sigma} \y O \z \> \equiv \frac{2}{\sqrt{g \x}} \frac{2}{\sqrt{g \y}}\frac{\delta}{\delta g^{\mu\nu}\x}\frac{\delta}{\delta g^{\rho\sigma}\y}\frac{\delta}{\delta J \z}\mathcal{Z}\left[g , J_i \right]\Big|_{g=\delta,J_i=0}.
\e
The only term in the derivative expansion of the generating functional that could be responsible for a momentum$^3$ term in $\< T_{\mu\nu}  T_{\rho\sigma}  O  \> $ is the gravitational Chern-Simons term for the spin connection $\omega_\mu$:
\b
\frac{1}{192\pi}\int \!\! d^3 x \sqrt{g} J \epsilon^{\mu\nu\rho}{\rm Tr}\left(\omega_\mu \partial_\nu \omega_\rho + \frac{2}{3} \omega_\mu\omega_\nu\omega_\rho\right),
\e
which is  diffeomorphism invariant only for constant $J$. By the previous line of reasoning, this term cannot be present in the derivative expansion of the generating functional. 
%Analogously to the case of the $U(1)$ current, the non-renormalization of the stress tensor's two point point function stems from diffeomorphism non-invariance of the Chern-Simons term for 
%This result can be equivalently stated in terms of the vacuum polarization tensor of a background graviton, thus making a connection with the $U(1)$ current discussed in the previous section. The observation that the relevant term in the generating functional  is not gauge invariant in the presence of the source $J$, translates to a similar statement regarding  diffeomorphism non-invariance.

\section{Discussion}
\label{section:Discussion}

In this work, we re-derived the Coleman-Hill theorem as an application of a more general argument, which was then used to prove the non-renormalization of the parity odd part of the energy-momentum two-point function at zero momentum. Moreover, 
as we do not assume minimal coupling, these conclusions hold also for nontrivially improved
currents. We expect that an analogous argument regarding the non-renormalization of the parity odd parts of various currents in any number of odd dimensions could be proven using the type of arguments presented here.  

%It should be stressed that these results are perturbative. %However, since the entire perturbative expansion is treated as a whole, in cases where the perturbation series is strongly asymptotic they may hold non-perturbatively as well. 

A subtle issue regarding the conditions of the theorem is worth some attention. The two main ingredients of the proof %, other than general properties of QFT,
seem to be the existence of a mass gap and a Lorentz invariant quantization; the first %ensures us of 
provides the analyticity of correlation functions at zero momentum and the second allows us to trust the properties of their tensor structure. At face value, the theorem should also apply when the gauge group is non-Abelian. However, even though an asymptotically free non-Abelian theory (with massive quarks) contains no massless asymptotic states, it does not respect our restrictions due to the need to add (massless) ghosts during quantization. One could attempt to decouple the ghosts by using an axial gauge, but then Lorentz invariance will not be manifest. 
The discussion presented here follows the spurion analysis of \citep{Closset2012}, where it was pointed out that $\<T_{\mu\nu} T_{\rho\sigma}\>$ does in fact receive higher loop corrections when the gauge group is non-Abelian, due to the topological structure of the theory \citep{Witten1989}. This issue is currently investigated by a perturbative analysis.

\acknowledgments

I would like to thank Guy Gur-Ari, Ran Yacoby and especially Shmuel Elitzur and Adam Schwimmer for interesting conversations and useful comments.
I am grateful for illuminating discussions with Zohar Komargodski, who also suggested this project. 
This work was supported
in part by the BSF \textendash{} American-Israel Bi-National Science
Foundation, and by a center of excellence supported by the Israel
Science Foundation~(grant~1665/10).
\newpage

\bibliographystyle{unsrt}
\bibliography{/Users/tomershacham/Dropbox/library}
%\bibliography{/home/shacham/Dropbox/library}
%\bibliography{library}
\end{document}